\documentclass[a4paper,11pt]{article}
\pdfoutput=1 
\usepackage{jcappub}

\usepackage{float}
\usepackage{color}
\usepackage{xcolor}
\usepackage{booktabs}
\usepackage{amsmath}
\usepackage{wrapfig}

\usepackage{graphicx}
\usepackage{bm}
\usepackage{epstopdf}

\usepackage{subfigure}
\usepackage{amsmath}
\usepackage{amsfonts}
\usepackage{amssymb}
\usepackage{ulem} 
\usepackage{graphicx}
\usepackage{natbib}

\title{\boldmath Estimating the Neutrino Flux from Choked Gamma-Ray Bursts}

\author[1]{Michela Fasano,}
\author[1,2]{Silvia Celli,}
\author[3,4]{Dafne Guetta,}
\author[1,2]{Antonio Capone}
\author[1,2]{Angela Zegarelli}
\author[1,2]{and Irene Di Palma}

\affiliation[1]{Dipartimento di Fisica dell'Universit\`a La Sapienza, P. le Aldo Moro 2, I-00185 Rome, Italy}
\affiliation[2]{Istituto Nazionale di Fisica Nucleare, Sezione di Roma, P. le Aldo Moro 2, I-00185 Rome, Italy}
\affiliation[3]{ORT-Braude College, Carmiel, Israel}
\affiliation[4]{Ariel University, Ariel, Israel}
\emailAdd{michelafasano5@gmail.com}
\emailAdd{silvia.celli@roma1.infn.it}
\emailAdd{dafneguetta@braude.ac.il}
\emailAdd{antonio.capone@roma1.infn.it}
\emailAdd{angela.zegarelli@roma1.infn.it}
\emailAdd{irene.dipalma@roma1.infn.it}

\abstract{The strong constraints from the Fermi-LAT data on the isotropic gamma-ray background suggest that the neutrinos observed by IceCube might possibly come from sources that are hidden to gamma-ray observations. A possibility recently discussed in the literature is that neutrinos may come from jets of collapsing massive stars which fail to break out of the stellar envelope, and for this reason they are known as \textit{choked jets}, or choked Gamma-Ray Bursts (GRBs). In this paper, we estimate the neutrino flux and spectrum expected from these sources, focusing on Type II SNe. We perform detailed calculations of $p\gamma$ interactions, accounting for all the neutrino production channels and scattering angles. We provide predictions of expected event rates for operating neutrino telescopes, such as ANTARES and IceCube, as well as for the future generation telescope KM3NeT. We find that for GRB energies channeled into protons spanning between $\sim 10^{51}-10^{53}$~erg, choked GRBs may substantially contribute to the observed astrophysical neutrino flux, if their local rate is $\sim 80 - 1$~Gpc$^{-3}$yr$^{-1}$ respectively.
}

\notoc

\begin{document}
\maketitle
\flushbottom
\newpage

\section{\textbf{Introduction}}
The diffuse flux of high-energy neutrinos observed by the IceCube experiment \cite{icecube2013} has not shown yet the presence of a significant event clustering \cite{icecubePS}. In other words, the angular distribution of the events collected so far does not highlight any specific location of the sky to search for correlated counterparts, e.g. in the electromagnetic domain or in the cosmic-ray anisotropy. This might indicate that the majority of neutrino sources is of extra-galactic origin.

The possible production mechanisms of neutrinos are respectively proton-proton (pp) collisions and photo-meson (p$\gamma$) interactions. In both cases, neutrinos would emerge from both the direct decay of charged pions as well as from muon decay, while gamma rays result from the neutral pion sudden decay. As such, within hadronic sources, a similar amount of high-energy electromagnetic radiation accompanies the production of neutrinos.

However, in the presence of intense radiation fields intrinsic to the source, the emitted gamma rays are likely to be absorbed in the pair production process (see e.g.~\cite{celli2017}). In addition, during their propagation towards Earth, multi-TeV gamma rays can initiate electromagnetic cascades on both the extra-galactic background light (EBL) and the cosmic microwave background (CMB), and be reprocessed towards lower energies. By comparing the isotropic diffuse gamma-ray background (IGRB), observed by Fermi \cite{fermi2015}, with the astrophysical neutrino flux, a strong tension among the two was remarked \citep{murase2015}, in the hypothesis that the pp interaction in transparent gamma-ray sources is originating both fluxes. This constraint motivates the alternative study of p$\gamma$ sources, opaque to high-energy gamma rays, as main neutrino sources.\\Choked GRBs belong to such a population \cite{meszaros2001}. These GRBs are believed to be originated from core collapse of massive stars resulting in a relativistic fireball jet,  which is unable to break through the stellar envelope. A choked jet deposits all its energy in a cocoon, that eventually breaks out from the star releasing energetic material at sub-relativistic velocities. Such fast moving material has a unique signature that can be detected in early time supernova spectra (see \cite{piran2017,izzo2019,xue2019}). Neutrino emissions from choked GRBs have hence attracted much attention \citep{denton2018,he2018,esmaili2018,guetta2019}, because of the possibility of explaining the IceCube diffuse flux without incurring into inconsistencies with the IGRB.

Estimates of neutrino fluxes from astrophysical sources as a result of the $p\gamma$ interaction have been performed in the past both with analytical \cite{atoyan2003,kelner2008} and numerical calculations \cite{lucarelli,murase2006,murase2008,hummer2010,murase2013,senno2016}, but no simulation for choked GRBs sources has been attempted so far. To study the expected neutrino flux on Earth from such $\nu$ bright (while $\gamma$ dark) sources, we have developed a Monte Carlo (MC) code to simulate the particles kinematics and interactions inside the jet, allowing for a detailed description of the micro-physics involved in the photo-meson process. Once the particle spectra at the source are obtained, our code calculates propagation through cosmic distances from the source to Earth, thus allowing a reliable estimate of the number of neutrino events expected in current and future neutrino telescopes.

The paper is organized as follows: Section \ref{sec:1} briefly describes the jet dynamics inside the source in order to characterize the interacting particle spectra. Section \ref{sec:2} presents an outline of the main structure of our simulation for the p$\gamma$ interaction. Calculations leading to the estimate of neutrino events in telescopes from a single source are contained in Section \ref{sec:4}. Section \ref{sec:5} describes the prediction for the diffuse neutrino spectrum coming from choked sources. Section \ref{sec:6} summarizes and discusses the results obtained. Moreover, Appendix \ref{app:a} illustrates the detailed kinematic calculations adopted in the code. 

\section{Choked jets and primary particle spectra}
\label{sec:1}
At the end of their lives, massive stars collapse into a compact object, and it is widely believed that aforesaid phenomena can produce energetic jets \citep{macfadyen1999,aloy2000}. GRBs are usually associated to type Ic SNe \cite{woosley}, which lack of the hydrogen envelope and have lost part of the helium envelope as well. However, much less is known concerning choked GRBs. There have been many investigations in the literature about their progenitor nature. The case of Type Ib/c SNe, corresponding to the scenario of failed GRBs, would imply a duration of the central engine less than 100~s, which is typical for long GRBs \cite{meszaros2001, razzaque2005, murase2013, senno2016, sobacchi2017}.
Several authors have also considered the possibility that  jets may be choked in the stellar envelope of red/blue supergiant stars, hence associated with Type II SNe (see e.g. \cite{he2018}, but also \cite{meszaros2001} consider  that this model is possible). In this case the duration of the central engine could be longer than for typical long GRBs \cite{xiao, kumar2008}. Note that this model may be problematic as most of the ordinary SNe may not have rapid rotation (see e.g. \cite{janka}).  While the details of the progenitor model remain uncertain, preliminary calculations suggest that a relativistic jet can be launched along the progenitor rotation axis \citep{meszaros2001}.

We consider a jet propagating inside the progenitor source and define $t_{\rm{jet}}$ as the jet lifetime. Such a parameter strongly depends on the physical process which powers the jet. We note that long GRBs are typically observed with durations shorter than 100~s. However, choked GRBs might behave significantly differently from standard ones. For instance, $10^5-10^6$~s lifetimes are rather expected if the jet is powered by accretion onto a central black hole or by the spin down luminosity of a magnetar \cite{he2018}. Hence $t_{\rm{jet}}$ should be considered as a free parameter of the model, possibly constrained by e.g. neutrino observations. \\
While the jet is making its way through the star, it can be slowed down in a termination shock to a Lorentz factor much lower than its original value. If the jet crossing time $t_{\rm{cross}}$ is longer than $t_{\rm{jet}}$, then the jet will result choked inside the stellar envelope, namely it will not be able to break through the star surface \citep{macfadyen1999}. In other words, when the central engine powers the jet for a shorter time than the one required to travel across the star radius, the jet is stalled inside the source. Since the jet crossing time reads as
\begin{equation}
t_{\rm cros} \simeq 1.1 \times 10^{4} \, \rm{s}\, R^2_{13.5} \, L_{\mathrm{iso, 50}}^{-1/2}\, \rho_{\rm H,-7}^{1/2} \, ,
\end{equation}
it is clear that extended stellar envelopes of radius $R$ are required to choke a luminous jet. Here (and elsewhere) $A_n$ denotes $A/10^n$, e.g. $L_{\mathrm{iso, 50}}^{1/8}=(L_{\rm{iso}}/10^{50})^{1/8}$, and cgs units are used for normalization. This might be the case of type II SN progenitors, namely core collapse of red supergiant stars, possessing thick stellar envelopes (see e.g. \cite{meszaros2001} for a detailed description of this source class). Such stars are surrounded by helium and hydrogen envelopes, which have radii respectively equal to $r_{\rm{He}}\sim 10^{11}\,\rm{cm}$ and $r_{\rm{H}}\gtrsim 10^{13}\,\rm{cm}$. Note that the source density in the hydrogen envelope is low enough, $\rho_{\rm{H}}\sim 10^{-7}\,\rm{g/cm^{3}}$, as to allow for particle acceleration to relativistic velocities, possibly followed by photomeson interactions beyond the helium radius. Alternatively, choked GRBs might originate from low luminous jets propagating into thinner stellar structures, as those belonging to massive stars which have lost a substantial fraction of their external envelopes through their final evolutionary stages. While the origin of choked jets remains so far elusive, in this work we will consider type II SN as a reference scenario, by adopting within our simulation the following benchmark values characterizing the jet features: the Lorentz factor $\Gamma=100$, the lifetime $t_{\rm{jet}}=10^3\,\rm{s}$ and the isotropic luminosity $L_{\rm{iso}}=10^{50}\,\rm{erg/s}$, where the two latter values provide a GRB isotropic energy given by $E_{\rm{iso}}=t_{\rm{jet}}\times L_{\rm{iso}}=10^{53}\,\rm{erg}$. We plan to investigate the type Ib/c scenario in a future work. \\

Internal shocks  (IS) are expected to develop in the jet interior, inward of the termination shock, at a distance from the inner engine equal to
\begin{equation}
R_{\rm IS} \simeq 2 \Gamma^2 c \delta t \simeq 6 \times 10^{12}~{\rm cm} \, \Gamma_2^2 \, \delta t_{-2} \, ,
\end{equation}
where $\delta t$ represents the minimum variability timescale of the central source. Depending on the initial opening angle $\theta$ of the jet and the properties of the environment where it propagates, the jet can result in a conical shape or in a cylindrical one \cite{bromberg2011}. The former case, also known as collimated jet, is verified if the jet's head Lorentz factor $\Gamma_{\rm h} \sim 1$ is smaller than $\theta^{-1/3}$. A non relativistic jet head satisfies the above condition for e.g. $\theta=0.2$~rad, which we will consider in the following. In addition, in such a case, the location of the jet head at the end of the jet lifetime can be approximated with \cite{he2018}
\begin{equation}
\label{eq:rh}
R_{\rm h} \simeq 9 \times 10^{12}~{\rm cm} \, L_{\mathrm{iso, 50}}^{1/4}\, t_{\rm{jet,\,3}}^{1/2} \, \rho_{\rm H,-7}^{-1/4} \, , 
\end{equation}
In between internal shocks and the jet head position, the collimation shock (CS) separates the shocked jet from the unshocked one, and it is found at \cite{mizuta2013,murase2013}
\begin{equation}
R_{\rm CS} \simeq 8.2 \times 10^{12}~{\rm cm} \, t_{\rm jet,3}^{2/5} \, L_{\mathrm{iso, 50}}^{3/10} \, (\theta_{\rm j}/0.2)^{2/5} \, \rho_{\rm H,-7}^{-3/10} \, .
\end{equation}
While the collimation shock is most likely radiation dominated, this is not the case for internal shocks, that are instead collisionless, and can hence accelerate protons and electrons to $\gtrsim 10^5$ GeV \citep{meszaros2001}. In fact, internal shocks are such that the co-moving size of the upstream flow is smaller than the mean free path of thermal photons in the upstream flow \cite{murase2013}. This constraint translates into a conservative bound of
\begin{equation}
l_{\rm u} < l_{\rm dec} \longrightarrow \tau=n_{\rm p,IS} \sigma_{\rm T} (R_{\rm IS}/\Gamma) < 1 \,,
\end{equation}
which for our benchmark parameters is verified as
\begin{equation}
\tau = 0.013 \Gamma_2^{-3} L_{\mathrm{iso, 50}}^{3/4} \, t_{\rm{jet,\,3}}^{-1/2} \,  \rho_{\rm H,-7}^{1/4} < 1 \, .
\end{equation}

Hence efficient acceleration of protons is expected to be realized at internal shocks, as also indicated from the comparison among the acceleration time of protons $t_{\rm acc}$ and the characteristic dynamical time of the system, e.g. $t_{\rm jet}$. The former reads as \cite{he2018}
\begin{equation}
t_{\rm acc}(E_p) = \phi \frac{E_p}{q_e Bc} \simeq 0.05 \, \rm{s}\, \phi_1 \, \epsilon^{-1/2}_{b,-1} \, E_{p,15} \, L_{\mathrm{iso, 50}}^{-1/4} \, \Gamma_2 \, t_{\rm{jet,\,3}}^{1/2} \, \rho_{\rm H,-7}^{-1/4} \, ,
\end{equation}
where $\phi$ is the number of gyro-radii required to e-fold the particle energy (here given in eV units). This estimate shows that multi-PeV protons can actually be accelerated at internal shocks within a time much shorter than the jet lifetime. Note that the strength of the magnetic field in the internal shock reads as 
\begin{equation}
B \simeq 2.9 \times 10^4 \, \rm{G} \, \epsilon^{-1/2}_{b,-1} \, \Gamma_2^{-1} \, L_{\mathrm{iso, 50}}^{1/4} \, t_{\rm{jet,\,3}}^{-1/2} \, \rho_{\rm H,-7}^{1/4} \, .
\end{equation}


Though little is known on the acceleration process in GRBs, we assume a first order Fermi process for primary charged particles \cite{fermi49}, producing a characteristic power-law differential energy distribution as:
\begin{equation}
\label{eq:2_1}
\frac{dN_p}{dE_p}= k_{\mathrm{MC}}\,E_p^{-2} \, ,
\end{equation}
where $k_{\mathrm{MC}}$ is a normalization factor which will be discussed in detail in Section \ref{sec:4}. Electrons are expected to lose all their energy into synchrotron radiation. However, because of the large optical thickness due to Thomson scattering, these photons thermalize in the reverse shock region (RS) to a temperature $T_\gamma$ such that \citep{razzaque2005}:
\begin{equation}
\label{eq:1_1}
kT_{\gamma}\simeq 313\,\rm{eV}\, \,L_{\rm{iso,\,50}}^{1/8}\,\epsilon_{e,\,-1}^{1/4} \,t_{\rm{jet,\,3}}^{-1/4}\, \rho_{\rm H,-7}^{1/8} \, ,
\end{equation}
$k$ being the Boltzmann constant, and $\epsilon_{e}$ the fraction of $E_{\rm{iso}}$ deposited into electrons. For the fraction of energy channeled into electrons and magnetic field we consider the equipartition condition $\epsilon_{\rm e}=\epsilon_{\rm b}=0.1\,E_{\rm{iso}}$.\\

Inside the RS, the photon spectrum is well represented by a blackbody distribution, reading as:
\begin{equation}
\label{eq:2_2}
\frac{dN_\gamma}{dE_\gamma}= \frac{8}{19(kT_{\gamma})^3} \frac{E_\gamma^2}{e^{\frac{E_\gamma}{kT_{\gamma}}} -1} \,,
\end{equation}
with a peak in the energy distribution given by the Wien's displacement law
\begin{equation}
\label{eq:1_2}
{E}^{\mathrm{\,max}}_{\gamma,\, \mathrm{RS}}\simeq 2.82\,kT_{\gamma}\sim 881\,\rm{eV}\,,
\end{equation}
and a number density following the Planck distribution
\begin{equation}
n_{\gamma,{\rm RS}}= \int_0^{\infty} \frac{dN_\gamma}{dE_\gamma} dE_\gamma= 16 \pi \xi(3) \left( \frac{kT_\gamma}{hc} \right)^3 \simeq 19.2 \pi \left( \frac{kT_\gamma}{hc} \right)^3\, ,
\end{equation}
where $\xi(3) \simeq 1.202$ is the Riemann zeta function. 
A fraction of thermal photons will then escape in the IS, since it is optically thin, where their peak energy becomes:
\begin{equation}
\label{eq:1_3}
{E}^{\mathrm{\,max}}_{\gamma,\, \mathrm{IS}}= \Gamma_{\rm{IR}} {E}^{\mathrm{\,max}}_{\gamma,\, \mathrm{RS}} \sim 88.1\, \mathrm{keV},
\end{equation}
where $\Gamma_{\rm{IR}}\sim\Gamma$ (if $\Gamma\gg 1$ \cite{he2018}) is the Lorentz factor of the IS with respect to the RS. 
The fraction $f_{\rm esc}$ of photons getting in the IS frame by Thomson scattering from the location $R_{\rm h}$ of the jet's head can be computed following \cite{murase2013}, namely
\begin{equation}
f_{\rm esc}= \frac{1}{\tau_{\rm h}}=\frac{1}{n_{\rm h} R_{\rm h} \sigma_{\rm T}} \, ,
\end{equation}
where $n_{\rm h}$ represents the number density of electrons in the shocked head region, which is related to the number density $n_{\rm j}$ of the unshocked jet through the compression ratio at the shock, i.e. $n_{\rm h}=4\Gamma n_{\rm j}$ for a strong shock and $n_{\rm j}=L_{\rm iso}/(4\pi c \Gamma^2 R_{\rm h}^2 m_{\rm p}c^2)$. Considering the jet head radius from Eq.~\eqref{eq:rh} we can derive
\begin{equation}
f_{\rm esc}= 1.9 \times 10^{-3} \, \Gamma_2 L_{\mathrm{iso, 50}}^{-3/4}\, t_{\rm{jet,\,3}}^{1/2} \, \rho_{\rm H,-7}^{-1/4} \, .
\end{equation}
Hence, the photon density in the internal shock frame reads as
\begin{equation}
\label{eq:ngammaIS}
n_{\gamma,{\rm IS}} = \Gamma n_{\gamma,{\rm RS}} f_{\rm esc} \simeq 1.9 \times 10^{20}\, \rm{cm^{-3}}\, \Gamma_2^2 \, L_{\mathrm{iso, 50}}^{-3/8}\, t_{\rm{jet,\,3}}^{-1/4}\,\epsilon_{e,\,-1}^{3/4} \, \rho_{\rm H,-7}^{1/8} \, .
\end{equation}

\section{\textbf{Photomeson interaction and neutrino production}}
\label{sec:2}
The production of charged and neutral mesons inside a jet can take place through the resonant production of a $\Delta^+$ by highly relativistic protons that interact with photons of the ambient radiation field, namely the thermal photons from the jet head. The code simulates the photo-meson interaction for each pairs of protons and photons which satisfy the threshold condition. The secondary particles emerging from different interaction channels are also simulated. Their energy losses in the form of synchrotron radiation, inverse Compton scattering off the dense thermal radiation field, synchrotron self Compton and adiabatic losses are accounted for. As we will show in the following, we identify synchrotron and adiabatic losses as the main channel for muon losses, while negligible losses affect charged pions. 
In order to account for possible energy losses of protons due to photomeson interactions, we compared the proton acceleration time with the $p\gamma$ loss time, the latter reading as
\begin{equation}
t_{\rm cool, p\gamma} = \frac{1}{n_{\gamma, \rm IS} c \sigma_{p\gamma} K_{p\gamma}} \simeq 0.0029  \, \rm{s} \, \epsilon^{-3/4}_{e,-1} \, \Gamma_2^{-2} \, L_{\mathrm{iso, 50}}^{3/8} \, t_{\rm{jet,\,3}}^{1/4} \, \rho_{\rm H,-7}^{-1/8} \, ,
\end{equation}
for a cross section $\sigma_{p\gamma} =1 \times 10^{-28}$~cm$^2$ and inelasticity $K_{p\gamma}=0.6$ \cite{atoyan2001}. From such a comparison, we obtain a proton cut-off of 
\begin{equation}
E^{\rm IS}_{\rm cut,15} \simeq 0.066 \, \phi^{-1}_1 \, \epsilon^{1/2}_{b,-1} \, \epsilon^{-3/4}_{e,-1} \, L_{\mathrm{iso, 50}}^{5/8} \, \Gamma^{-3}_2 \, t_{\rm{jet,\,3}}^{-1/4} \, \rho_{\rm H,-7}^{1/8} \,,
\end{equation}
where again the proton energy is given in eV units, namely a cut-off at $E_{\rm cut}=66$~TeV arises in the IS frame. This affects the differential energy spectrum of protons, which is modified for $E_{\rm p}>E_{\mathrm{cut}}$ with respect to the simple power law of Fermi acceleration given in Eq.~\eqref{eq:2_1}, as
\begin{equation}
\label{eq:fermiMod}
\frac{dN_{\rm p}}{dE_{\rm p}}= k_{\mathrm{MC}}\,E_p^{-2}\,\, e^{-\frac{(E_{\rm p}-E_{\mathrm{cut}})}{2E_{\mathrm{cut}}}} \, .
\end{equation}
\noindent
The kinematics of particles emerging from the $\Delta^+$ production is evaluated in the center of mass rest frame, and then a Lorentz boost is applied to transform them to the IS frame through the center of mass Lorentz factor $\gamma^*$. All the center of mass quantities are indicated with an asterisk $*$. Ultimately, the observed energies at the source are obtained by boosting the particles by the jet Lorentz factor $\Gamma$. In the center of mass frame we use natural units where $\hbar $ = c = 1. Moreover, we indicate with $\beta_a$ the speed of the generic particle $a$ ($\gamma_a=1/\sqrt{1-\beta_a^2}$), and with $E_a$, $p_a$, $m_a$ its energy, momentum and mass. For detailed kinematic calculations see Appendix~\ref{app:a}.

Once a proton and a photon are extracted, in order to proceed with the interaction a condition must be met, i.e. the production of the $\Delta^+$ resonance, given by the threshold requirement (see Eq.~\eqref{eq:2_4}). In the IS, the condition for a successful interaction can be written as:
\begin{equation}
\nonumber
(m_p+m_{\pi})^2=({E}^{\rm{IS}}_{p}+{E}^{\rm{IS}}_{\gamma})^2-({\vec{p}}^{\,\rm{IS}}_{p}+{\vec{p}}^{\,\rm{IS}}_{\gamma})^2
\end{equation}
\begin{equation}
\nonumber
m_p^2+m_{\pi}^2+2 m_p m_{\pi}=m_p^2+2{E}^{\rm{IS}}_{p}{E}^{\rm{IS}}_{\gamma}(1-\cos\theta)
\end{equation}
\begin{equation}
\label{eq:2_4}
{E}^{\rm{IS}}_{p} \ge \frac{{m_{\pi}}^2+2m_p m_{\pi}}{2{E}^{\rm{IS}}_{\gamma} (1-\cos\theta)} \, ,
\end{equation}
where $\theta$ indicates the angle between protons and photons in the IS. When the threshold requirement in Eq.~\eqref{eq:2_4} is not satisfied, a new photon is extracted.
In fact, as a consequence of the extremely high number density of thermal photons (see Eq.~\eqref{eq:ngammaIS}), all generated protons will eventually interact.
In order to compute the proton path inside the thermal photon gas, it is necessary to invert the path probability distribution:
\begin{equation}
\label{eq:2_7}
\rm{dP(x_{\rm{p}})}= \frac{1}{\lambda_{p\gamma}}\ e^{-x_{\rm{p}}/\lambda_{p\gamma}}\ \rm{dx_{\rm{p}}} \, , 
\end{equation}
where $\lambda_{p\gamma} = 1/(\sigma_{p\gamma} n_{\gamma})$ is the interaction length. The proton path obtained through the inversion of Eq.~\eqref{eq:2_7} peaks at $\sim 10^7$~cm for $\sigma_{p\gamma}=550\,\mu \rm{b}$ (the $\Delta^+$ resonance peak), while for the out of resonance region value of $\sigma_{p\gamma}$, namely $\sim 120\,\mu \rm{b}$, the proton path peaks at $\sim 10^8$ cm. This means that a proton will interact, on average, after travelling $10^{7.5}$~cm inside the source, which has an external hydrogen envelope radius of ~$10^{13}$~cm. Thus, we can safely assume that all protons interact inside the source. In other words, choked jets behave as pure calorimeters with respect to protons above the pion production thresholds, such that all their energy goes into secondary light particles, including gamma rays and neutrinos. This result is consistent with findings from other authors as well (see e.g. \cite{senno2016}). The high-energy radiation emitted is expected to be reprocessed at lower energies in pair production processes with existing jet radiation field. Hence observations of neutrinos would provide a unique constrain on the energy channeled into accelerated protons.

Inside the IS region, the following relation is also satisfied:
\begin{equation}
\label{eq:2_5}
\cos{\theta}=\frac{{-p}^{\rm{IS}}_L}{\sqrt{({p}^{\rm{IS}}_L)^2+({p}^{\rm{IS}}_T)^2}} \, ,
\end{equation}
${p}^{\rm{IS}}_L$ and ${p}^{\rm{IS}}_T$ being, respectively, the longitudinal and transverse momenta of the photons in the IS.
Defining $\xi$ as the angle between protons and photons in the RS, we obtain the following algebraic equations for the photon momenta:
\begin{equation}
\nonumber
-{p}^{\rm{IS}}_L=-\Gamma_{\rm{IR}}\ E_{\gamma,\, \rm{RS}}\ (\cos\xi\ +\ \beta_{\rm{IR}})
\, ,
\end{equation}
\begin{equation}\nonumber
{p}^{\rm{IS}}_T={p}^{\rm{RS}}_T=\sqrt{E_{\gamma,\, \rm{RS}}^2\ (1-\cos^2\xi)}\,.
\end{equation}
For all simulated interactions, the cosine value obtained from Eq.~\eqref{eq:2_5} is $\sim 1$, meaning that the collisions are realized head-on.

Once the interaction occurs, the Monte-Carlo algorithm takes into account the various interaction channels. Within our simulation we considered three possible channels for the interaction, depending on the photon energy in the proton's rest frame, $\epsilon_r$:
\begin{equation}\nonumber
s^{\rm{IS}}=2{E}^{\rm{IS}}_{p}{E}^{\rm{IS}}_{\gamma}(1-\cos\theta)\sim 4{E}^{\rm{IS}}_{p}{E}^{\rm{IS}}_{\gamma}=m^2_p+2m_{p}{\epsilon}_{r} \, ,
\end{equation}
$s^{\rm{IS}}$ being the square of the invariant energy of the system in the IS frame, such that
\begin{equation}
\label{eq:32}
\epsilon_r=\frac{4{E}^{\rm{IS}}_{p}{E}^{\rm{IS}}_{\gamma}-m_p^2}{2m_p}\,.
\end{equation}

The p$\gamma$ cross section as a function of $\epsilon_r$ is characterized by a very peaked resonance region around the threshold, mainly due to the two reaction channels we are studying (namely the ones giving the $p + \pi^0$ and $n + \pi^+$ final states): in our simulation, the interaction probability $P_{\rm{int}}$ in this region is set equal to 1. Above the threshold, the cross section is dominated by multiple pion final states, and it becomes approximately constant for $\epsilon_r>$ 10 GeV. For a detailed characterization of the p$\gamma$ cross section see e.g. \citep{morejon2019}. Hence, multiple values
were used for $P_{\rm{int}}$ within the code, depending on the cross-section. 

The boundaries between each interaction channel in our simulation are given by the following numerical values:
\begin{itemize}
\item $0.2\, \mathrm{GeV}\le\epsilon_r<0.5\, \mathrm{GeV}$: this energy range corresponds to the peak of the cross section, where the interaction occurs with $P_{\rm{int}}=1$ and the $\Delta^+$ resonance is produced at rest: \begin{equation}
\label{eq:33}
p+\gamma \to \Delta^+ \to 
\begin{cases} n + \pi^+ \\ p + \pi^0
\end{cases};
\end{equation}
\item $0.5\, \mathrm{GeV}\le\epsilon_r<2\, \mathrm{GeV}$: such interval matches the secondary peak range of the cross section, when the $\Delta^+$ resonance is produced along with an additional pion ($\mathcal{N}$ is a nucleon, while both $\pi^{\rm{A}}$ and $\pi^{\rm{B}}$ can be neutral, positively or negatively charged pions):
\begin{align} 
p + \gamma \to  \pi^{\mathrm{A}} + &\Delta^+ \nonumber\\ &\to\,\,\mathcal{N} + \pi^{\mathrm{B}} \, .
\label{eq:34}
\end{align}
Here, the interaction occurs with two possible values of probability: $P_{\rm{int}}=0.6$ when $\epsilon_r\le1\,\rm{GeV}$, otherwise the interaction probability is set equal to 0.4; 
\item $2\, \mathrm{GeV}\le\epsilon_r\le100\, \mathrm{GeV}$: this energy window corresponds to the plateau in the cross section, in which the three types of pions are simultaneously produced. This range is often referred to as \textit{multipion channel}, where ``$\ldots$''  indicates other particles created in the interaction:
\begin{equation}\label{eq:35}
p+\gamma \to \pi^0 + \pi^+ +\pi^- + \ldots
\end{equation}
In this region $P_{\rm{int}}=0.3$.
\end{itemize}
The code selects how to progress the interaction based on which range $\epsilon_r$ occupies. The interested reader is referred to Appendix~\ref{app:a} for a more detailed discussion on branching ratios and kinematics in the three channels. \\
The pion decay products include leptons and photons:
\begin{align}
{\pi}^+ &\to \mu^+ + \nu_{\mu} \to e^+ +\nu_e+ \bar{\nu}_{\mu} + \nu_{\mu} \label{eq:36}\,
,\\
{\pi}^- &\to \mu^- + \bar{\nu}_{\mu} \to e^- +\bar{\nu}_e+\nu_{\mu} + \bar{\nu}_{\mu} \label{eq:37}\,
,\\
{\pi}^0 &\to \gamma + \gamma \label{eq:38}\,
.
\end{align}

The muon decay is taken into account, and characterized through the \textit{Michel parameters}, which describe the phase space distribution of leptonic decays of charged leptons. However, energy losses in the dense radiation fields or in the jet magnetic field might affect the spectrum of both pions and muons, possibly suppressing the flux of emerging neutrinos, if these losses develop on shorter timescales than the particle decay time. Concerning the muon, its lifetime at rest is $\tau_{\mu}\sim 2.2 \times 10^{-6}$~s, while for the charged pion its lifetime at rest is $\tau_{\pi^{\pm}}\sim 2.6 \times 10^{-8}$~s. Additionally, while the jet expands, part of the kinetic energy is lost in the form of heat, resulting into further energy losses for the particles as well. The characteristic timescale for adiabatic losses can be taken the same as the dynamical timescale of the system, namely
\begin{equation}
t_{\rm dyn} = \frac{R_{\rm IS}}{\Gamma c} = 2 \Gamma \delta t \simeq 2 \, \rm{s} \, \Gamma_2 \, \delta t_{-2} \, ,
\end{equation}
shorter than for instance the jet lifetime ($t_{\rm jet}=1000$~s). This scale has to be compared with the time taken to emit synchrotron radiation, or to energize the thermal or electron-emitted synchrotron photons to high-energy, through inverse Compton processes. \\
For synchrotron radiation, muons loose energy in a characteristic time of
\begin{equation}
\tau_{\mu}^{\rm{sync}}(E^{\rm IS}_\mu)=0.89\,\rm{s}\,\,E^{\rm{IS},\,-1}_{\mu,\,15} \epsilon_{b,\,-1}^{-1} \, \Gamma_2^2\, L_{\mathrm{iso, 50}}^{-1/2}\, t_{\rm{jet,\,3}} \, \rho_{\rm H,-7}^{-1/2} \, ,
\end{equation}
while charged pions of
\begin{equation}
\tau_{\pi^\pm}^{\rm{sync}}(E^{\rm IS}_{\pi^\pm})=2.60\,\rm{s}\,\,E^{\rm{IS},\,-1}_{\pi^\pm,\,15} \epsilon_{b,\,-1}^{-1} \, \Gamma_2^2\, L_{\mathrm{iso, 50}}^{-1/2}\, t_{\rm{jet,\,3}} \, \rho_{\rm H,-7}^{-1/2}.
\end{equation}
For inverse Compton we note that the IC interaction for muons and pions always proceed in the Klein Nishina regime, as these secondary particles emerge from the p$\gamma$, which only occurs at energies above threshold. We follow the approach from \cite{meszaros2001}, where the IC loss time of a pion in the IS frame off the thermal field reads as 
\begin{equation}
\label{eq:tauICp}
\tau^{\rm IC}_{\rm KN}(E_\pi^{\rm IS}) \simeq \frac{2}{m_e^2c^4} \frac{\Gamma^2 E_\pi^{\rm IS} \epsilon^{\rm RS}_\gamma}{c\sigma_{\rm T} n^\pi_\gamma} \, ,
\end{equation}
where $n^\pi_\gamma=2.9 \times 10^{17}$~cm$^{-3}$ with our benchmark values and $\sigma_{p\gamma}=5\times 10^{-28}$~cm$^2$.
Plugging numbers in, we obtain for charged pions a timescale of
\begin{equation}
\label{eq:tauP}
\tau^{\rm IC}_{\rm KN}(E_\pi^{\rm IS}) \simeq 1.2 \times 10^4~{\rm s} \left(\frac{E_\pi^{\rm IS}}{1~{\rm TeV}} \right) \left(\frac{\Gamma}{100} \right)^2 \left( \frac{\epsilon^{\rm RS}_\gamma}{0.881~{\rm keV}} \right) \left(\frac{n^{\pi}_\gamma}{2.9 \times 10^{17}~{\rm cm}^{-3}} \right)^{-1}  \, 
.
\end{equation}
\noindent
An analogous expression holds for the IC loss time of muons. Thus we can argue that the effect of energy losses due to inverse Compton processes do not affect the pions and muon decays. \\


Finally, for the evaluation of the loss time in SSC process of pions and muons off the synchrotron radiation field emitted by primary electrons, we simply note that it can always be neglected with respect to the synchrotron loss time of the same particles, because of the equipartition condition. To summarize the results obtained with the benchmark parameters we set, charged pions do not suffer any significant loss process within their lifetime, namely they always decay freely. In turn, muons with energies higher than $E_\mu^{\rm IS}\simeq 100$~TeV decay with a lower probability because they are affected by adiabatic losses, and finally above $E_\mu^{\rm IS}\simeq 500$~TeV by synchrotron losses. These conditions are applied to muons in all interaction channels, and the corresponding neutrino spectra obtained through the simulation depend on such a suppression. It should be noted, however, that the cut applied to protons at the level of $E_{\rm cut}\simeq 66$~TeV in the IS frame reflects into a cut on the pion and muon energies stronger than the cut due to their own adiabatic and synchrotron losses. \\

\begin{figure*}[h!]
    \centering
    \includegraphics[scale=0.25]{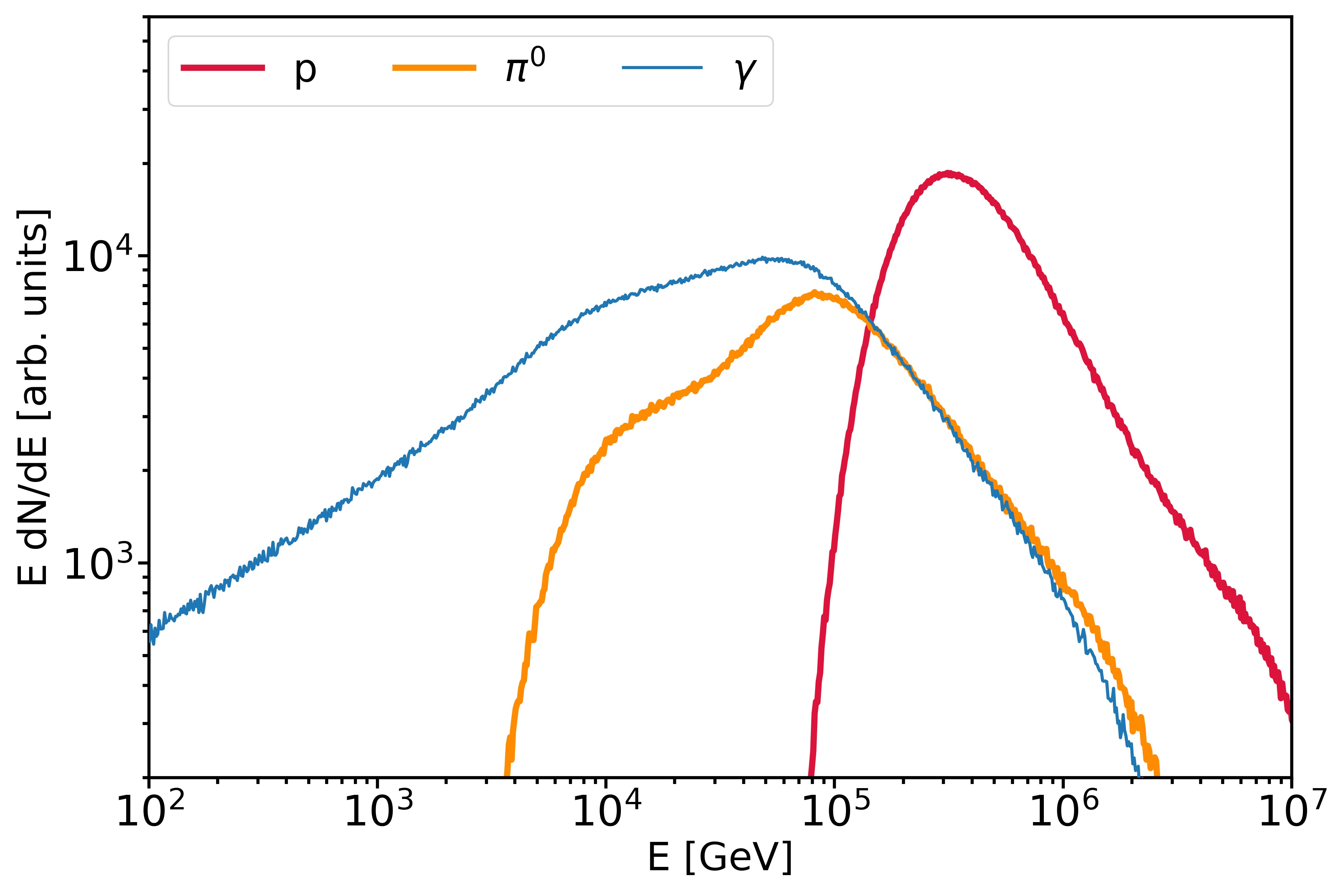}
    \includegraphics[scale=0.25]{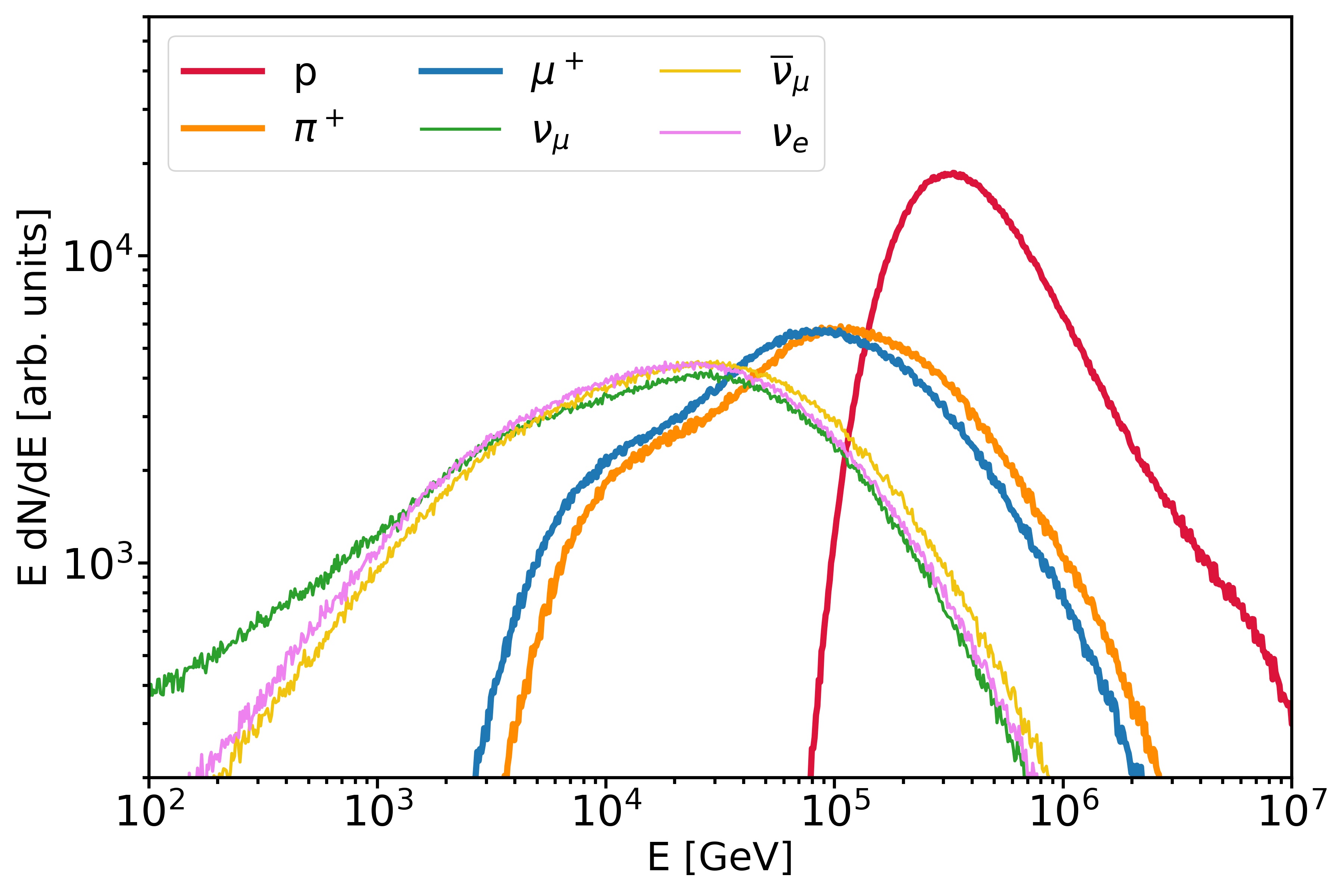}
    \includegraphics[scale=0.25]{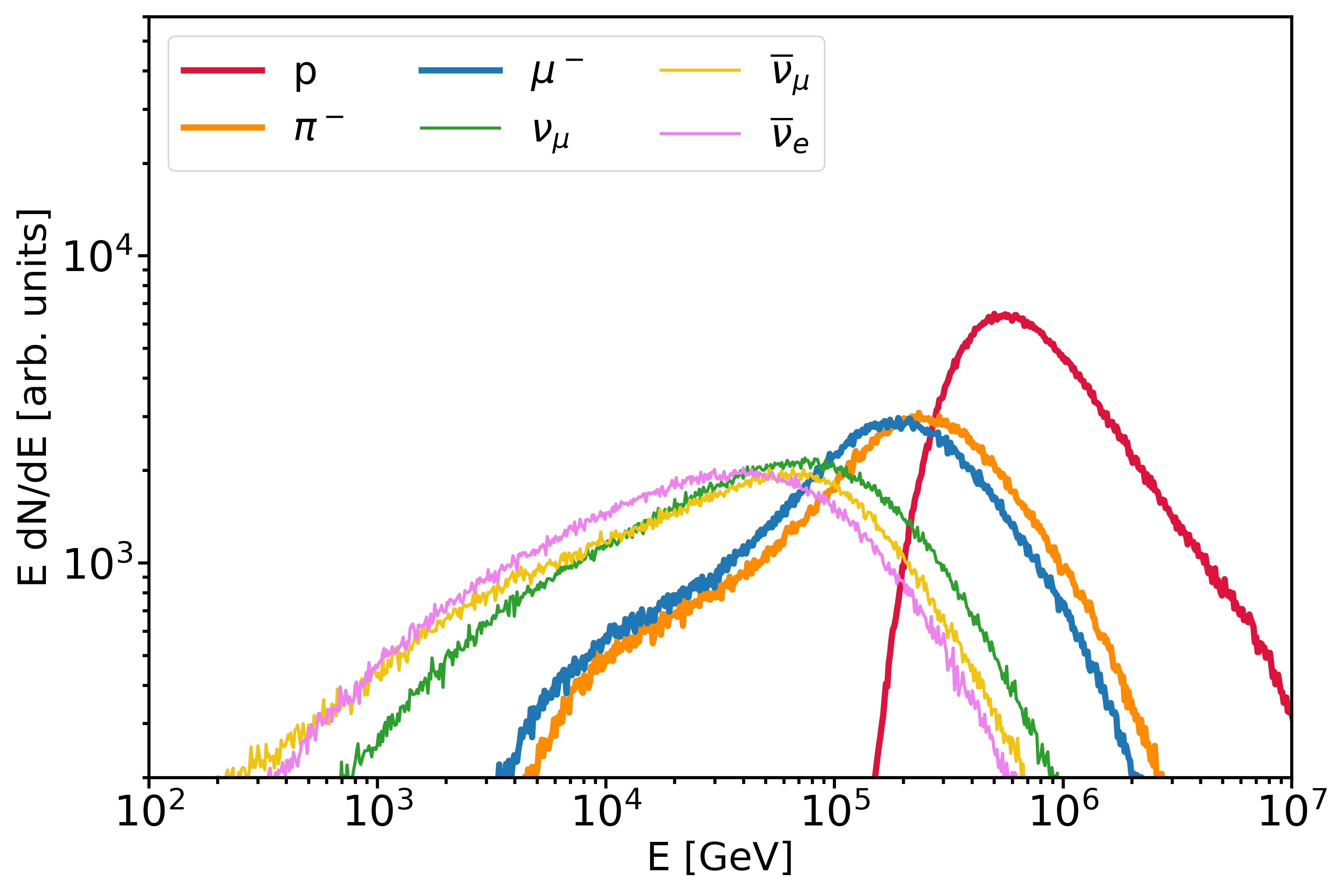}
    \caption{Particle spectra at the source arising from all interaction channels, as seen in the co-moving reference frame. \textbf{Top left}: $\pi^0$ channel; \textbf{Top right}: $\pi^+$ channel; \textbf{Bottom}: $\pi^-$ channel.}
    \label{fig:1}
\end{figure*}

The particle spectra at the source obtained through our MonteCarlo simulation are shown in Figure \ref{fig:1}. In the top left panel we present the energy spectrum of the interacting protons, as well as the secondary particles emerging from the $\Delta^+ \longrightarrow p+\pi^0$ decay channel. In this study we focus on neutrino production, hence we do not propagate the resulting gamma rays further in the jet, namely we neglect pair production processes of the high-energy photons with the thermal radiation field of the jet. We defer discussion about absorption of gamma rays in the chocked jet to a future work. In the top right panel, we show the secondary particles from all $\Delta^+$ decays that give rise to $\pi^+$, while in the bottom panel we show the spectrum of $\pi^-$ obtained from $\Delta^+$ decay in the multi-pion decay channel. Note that the latter is extremely similar to the $\pi^+$ channel, while it contains far less events due to the fact that this channel is not present at the resonance peak, where most of the particles are produced. These spectra are not normalized yet: as a consequence, the y axis is in arbitrary units.

\section{Number of events expected in neutrino telescopes}
\label{sec:4}
Transient phenomena such as GRBs are very interesting for a potential neutrino detection, since these analyses are almost background free given the well-defined space-time window provided by multi-messenger constraints. To compute the neutrino flux expected on Earth from a source at a cosmological distance, we scale the MC-obtained fluxes for three factors: the energy channeled into protons at the source, the dilution of the neutrino flux to Earth, and neutrino oscillations. 

The first factor comes from energetic considerations for the source: as mentioned in Section \ref{sec:1}, we considered a choked GRB with an isotropic energy release of $E_{\rm iso}= 10^{53}$~erg. We also assume that a comparable amount of energy $E_{\rm p,jet}$ is delivered to the protons accelerated inside the jet \citep{meszaros2001}. So, for the accelerated protons spectrum at the source in the IS frame we have:
\begin{equation}
\label{eq:4_1}
k_{\rm MC} \int_{E_{\rm min}}^{E_{\rm max}} E_{\rm p} \frac{dN_{\rm p}}{dE_{\rm p}} dE_{\rm p} = E_{\rm p,jet} \, ,
\end{equation} 
where $E_{\rm{min}} = 100$ GeV and $E_{\rm{max}} = 10^{8}$ GeV. Solving this integral with the acceleration spectrum of protons, namely the pure power law distribution given in Eq.~\eqref{eq:2_1}, 
leads us to an estimate of the scaling factor $k_{\mathrm{MC}}$,
which provides the exact fluence of secondary particles at the source in the IS frame, including gamma rays and neutrinos, normalized as to take into account the amount of energy channeled into accelerated protons. Moreover, particles travel through cosmological distances before arriving on Earth, such that their flux undergoes a dilution during the propagation. This factor can be calculated as $(1+z)^2/(4\pi d_{\rm L}(z)^2)$ \cite{baerwald,murase2007}, $d_{\rm L}(z)$ being the source luminosity distance. For the choked GRB in our model, we assume a redshift $z$ of $\sim 1$, from which we can evaluate $d_{\rm L}(z=1)=2.06\times 10^{26}$~m.

Neutrino telescopes adopt muon neutrino and muon antineutrino fluxes at Earth for astronomical studies, thanks to the good angular resolution obtained in this event sample by reconstructing the long track length of the induced muon. Neutrino flavor on Earth differs from the one at the source due to the effect of neutrino oscillations. Since GRBs are extra-galactic sources, we mediate neutrino oscillations over cosmic distances: starting from a flavor ratio of $\nu_e$~:~$\nu_{\mu}$~:~$\nu_{\tau}$~=~1~:~2~:~0 at the source (as the result of the $p\gamma$ interaction), one should expect a flavor ratio of 1~:~1~:~1 on Earth. The oscillation probabilities $P_{ff'}$ of neutrinos changing flavor ($f \to f'$) are given by \cite{villante}:
\begin{equation}
\label{eq:4_3}
P_{ff'}=\sum_{j} \left|U_{fj}^2\right| \left|U_{f'j}^2\right|,
\end{equation}
so that the oscillated spectrum for the lepton of flavor $f$ reads as:
\begin{equation}
\label{eq:4_4}
E_{\nu_f} \frac{dN_{\nu_f}}{dE_{\nu_f}} = P_{ff} \left( E_{\nu_f} \frac{dN_{\nu_f}}{dE_{\nu_f}} \right)^{\rm{\,source}} + P_{f^\prime f} \left( E_{\nu_{f^\prime}} \frac{dN_{\nu_{f^\prime}}}{dE_{\nu_{f^\prime}}}  \right)^{\rm{\,source}} \, ,
\end{equation}
where the right-hand side quantities represent the generic neutrino spectrum at the source in the co-moving frame, obtained through the MC simulation with the $\Gamma$-Lorentz boost from the IS frame. This formula is symmetric in the exchange $f \leftrightarrow f'$, and it is valid for neutrinos and antineutrinos.

\begin{figure}
    \centering
    \includegraphics[scale=0.25]{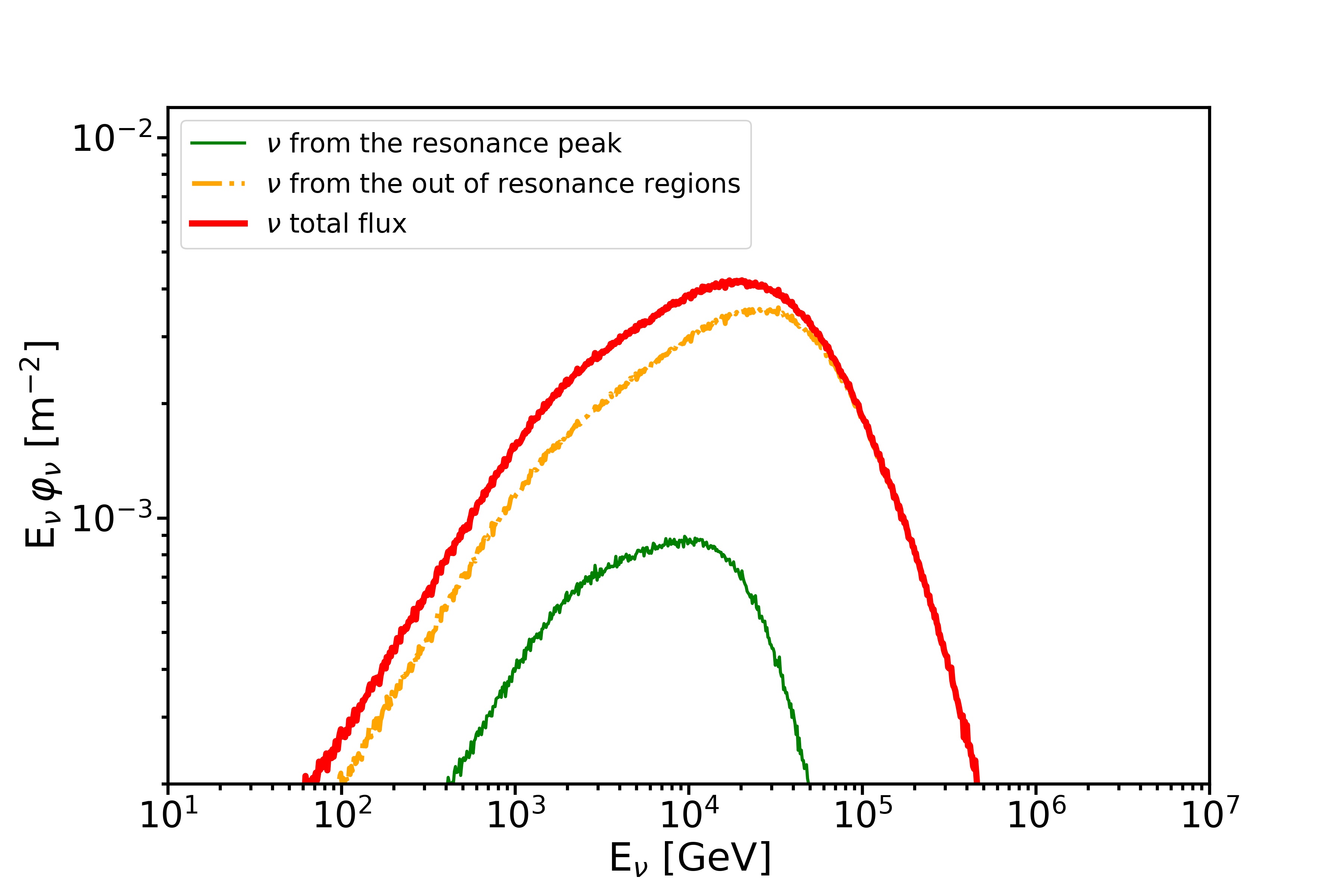}
    \includegraphics[scale=0.25]{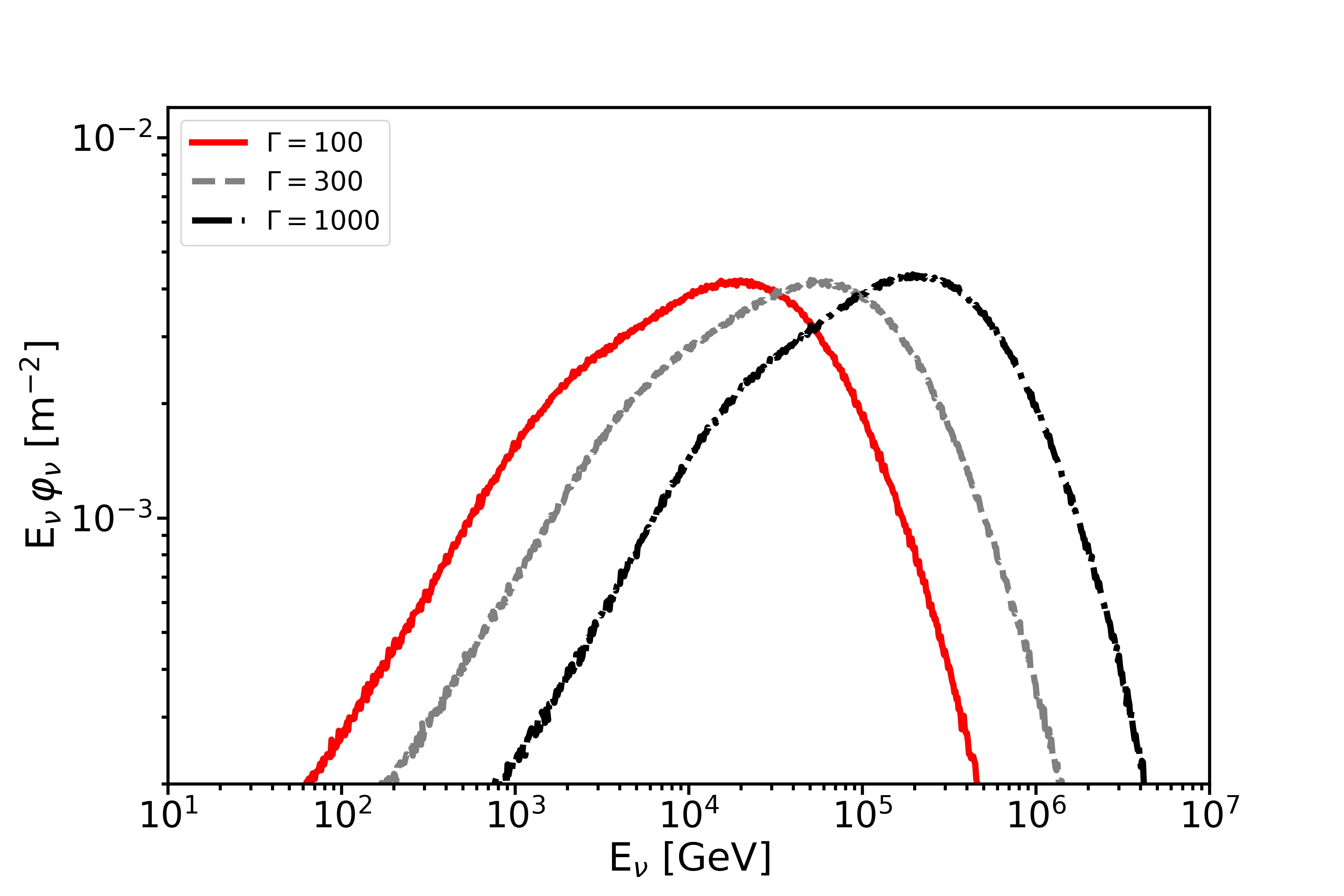}
    \caption{All-flavor time-integrated neutrino flux on Earth. \textbf{Left}: neutrinos from the resonance peak (green dashed line), from the region above the peak (orange dot-dashed line) and their sum (red line) are compared in this plot. \textbf{Right}: neutrinos from choked jets with different Lorentz factors.}
    \label{fig:2}
\end{figure}
\begin{figure}
    \centering
    \includegraphics[scale=0.33]{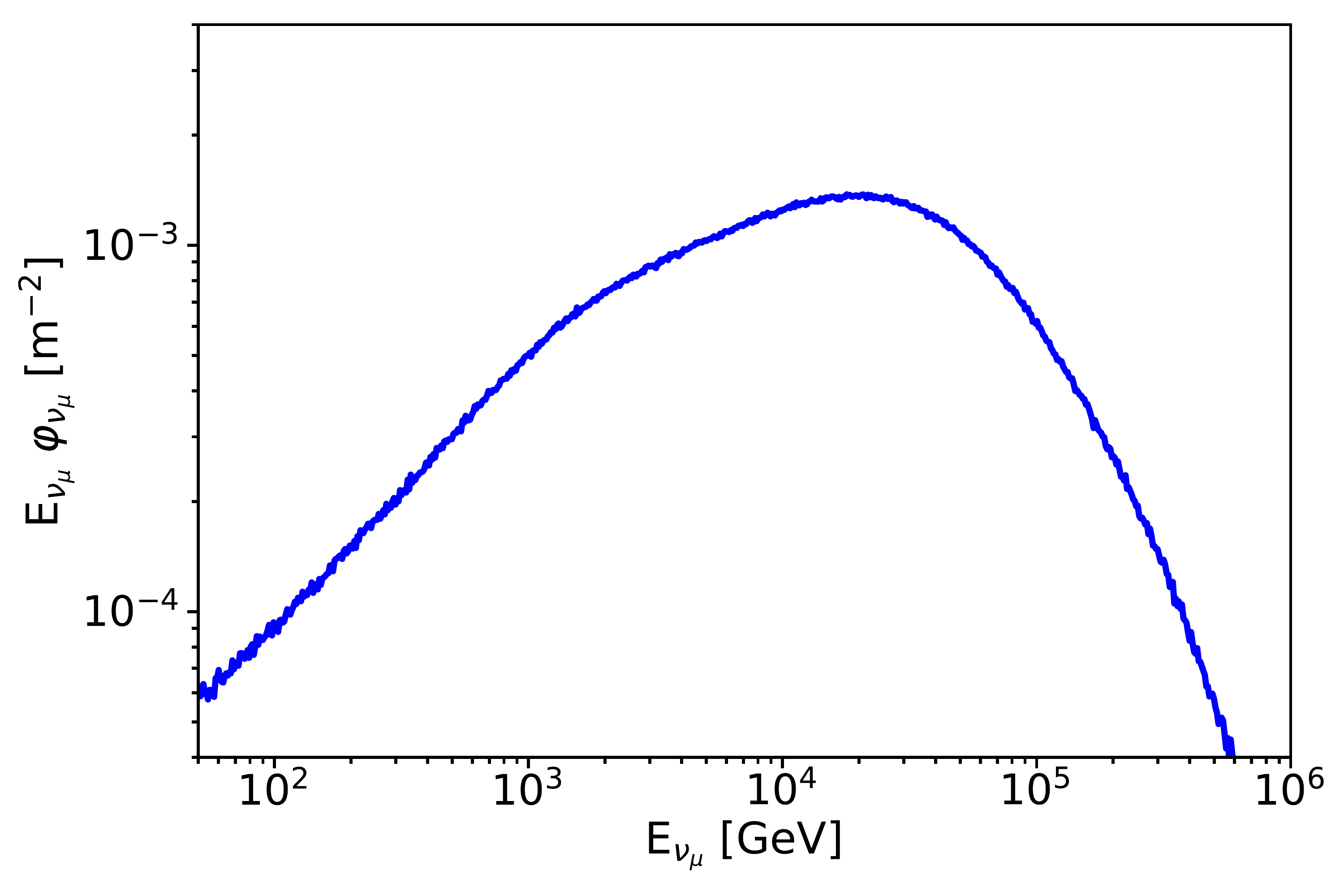}
    \caption{Muon neutrino and antineutrino time-integrated flux expected on Earth from a single source at z=1. Oscillations of neutrinos are here accounted for.}
    \label{fig:3}
\end{figure}

\noindent
Defining the total neutrino energy spectrum on Earth as
\begin{equation}
E_{\nu} \frac{dN_{\nu}}{dE_{\nu}} = E_{\nu_e} \frac{dN_{\nu_e}}{dE_{\nu_e}} + E_{\nu_\mu} \frac{dN_{\nu_\mu}}{dE_{\nu_\mu}} + E_{\nu_\tau} \frac{dN_{\nu_\tau}}{dE_{\nu_\tau}} \, ,
\end{equation}
the all-flavor time-integrated neutrino flux on Earth can be written as:
\begin{equation}
\label{eq:f_nu}
E_{\nu} \varphi_{\nu}^{\rm{Earth}}=k_{\mathrm{MC}}\,\frac{(1+z)^2}{4\pi\,d_{\rm L}(z)^2} E_{\nu} \frac{dN_{\nu}}{dE_{\nu}} \, .
\end{equation}
The all-flavor time-integrated neutrino flux on Earth can be seen in Figure \ref{fig:2}, where we also accounted for the redshift dependence of observed energy. In the left panel, neutrinos coming from the delta resonance peak region ($0.2\, \mathrm{GeV}\le\epsilon_r<0.5\, \mathrm{GeV}$) are compared with neutrinos coming form the energy range exceeding $0.5\, \mathrm{GeV}$. This result is particularly interesting because it indicates that focusing merely on the $\Delta^+$ resonance peak for neutrino production leaves out a non negligible number of secondary neutrinos coming from higher interaction channels. Furthermore, the right panel of Figure \ref{fig:2} compares the all-flavor neutrino spectra coming from choked jets with different Lorentz factors: the higher the $\Gamma$, the higher the neutrino energy peak. Conversely, the absolute normalization of the neutrino time-integrated flux $E_{\nu} \varphi_\nu$ (see Eq.~\eqref{eq:f_nu}) is independent of $\Gamma$, as $E_{\rm p,jet}$ was set in the IS frame.\\We are interested in estimating the number of muon neutrino events expected from an individual source in different neutrino telescopes, since these constitute the best channel of events for astronomical studies due to their excellent angular resolution. Hence, we proceed by oscillating neutrinos towards Earth, as to obtain the flux of muon neutrino flavor. The values for the oscillation probabilities $P_{\mu\mu}$ and $P_{e\mu}$ to be used in Eq.~\eqref{eq:4_4} are estimated using the latest results for the mixing angles, for neutrino mass hierarchy in the case of normal ordering \cite{mascaretti2019}. The muon neutrino time-integrated neutrino flux observed on Earth obtained within our simulation is shown in Figure \ref{fig:3}, by using the following set of benchmark parameters: $z=1$, $E_{\rm{p,\,jet}}=E_{\rm{iso}}=10^{53}$~erg, $L_{\rm{iso}}=10^{50}$~erg/s, $t_{\rm jet}=10^3$~s, and $\Gamma=100$.\\



\begin{figure}
    \centering
    \includegraphics[scale=0.35]{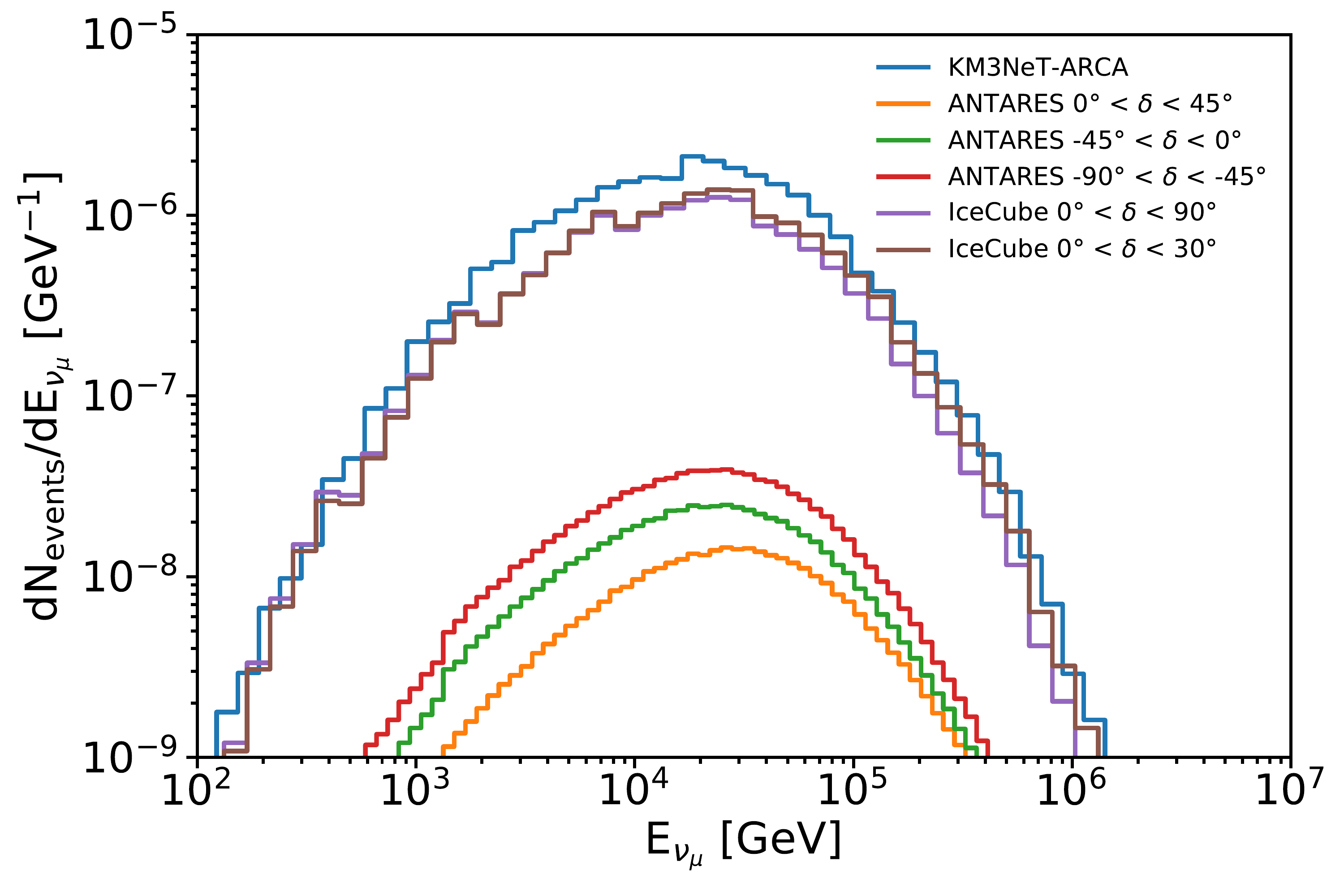}
    \caption{Number of expected events in ANTARES \cite{ANTARES}, KM3NeT-ARCA \cite{KM3} and IceCube \cite{IceCube} from a characteristic choked GRB located at different declinations.}
    \label{fig:4}
\end{figure}

Using the energy distribution of muon neutrinos thus obtained, an estimate of the number of cosmic neutrino events expected in neutrino telescopes can be performed. Convolving the detector effective area for neutrinos, $A_{\mathrm{eff}}^{\nu}(E_{\nu}, \delta)$, coming from declination $\delta$, with the differential number of neutrinos per surface element expected on Earth from the source, the number of events during the GRB event ($t_{\rm{jet}}=1000\,\rm{s}$) is obtained through the following relation:
\begin{equation}
\label{eq:4_7}
N_{\rm events}(\delta)= \int \frac{dN_{\rm events}}{dE_{\nu}} (E_{\nu}, \delta)\,dE_{\nu}=\int A_{\mathrm{eff}}^{\nu}(E_{\nu}, \delta)\,\Phi_{\nu}^{\rm{Earth}}(E_{\nu})\,dE_{\nu}
\,
.
\end{equation}
This evaluation has been performed for ANTARES \cite{ANTARES}, KM3NeT-ARCA \cite{KM3} and IceCube \cite{IceCube}. Note that the effective area adopted in this work for KM3NeT refers to the trigger level, being the only available information at the time of writing. In turn, the effective areas adopted for ANTARES and IceCube refer to the analysis level, as these result from detailed studies performed in the search for neutrino sources by the same Collaborations. The expected number of events per energy bin is shown as a function of the energy in Figure \ref{fig:4}, while the total number of events expected in each neutrino telescope during the choked GRB event is given in Table \ref{tab:1}. The calculations show that current generation detectors have non negligible chances to detect neutrinos from individual sources of this kind, in the assumption that choked GRBs can efficiently convert their kinetic energy into accelerated protons. In particular, if the rate of occurrence of such explosions in the Universe is high enough, it would be possible for the population of choked GRBs to contribute to the observed diffuse astrophysical neutrino flux, whose sources remain so far unidentified.

\begin{table}[h]
\centering
\begin{tabular}{lcc} 
\hline
Detector & $\delta$ & $N_{\mathrm{events}}$ \\ [0.5ex] 
\hline
ANTARES & $0^{\circ}<\delta<45^{\circ}$ & 2$\times 10^{-3}$ \\ 
& $-45^{\circ}<\delta<0^{\circ}$ & 3$\times 10^{-3}$ \\
& $-90^{\circ}<\delta<-45^{\circ}$ & 5$\times 10^{-3}$ \\
\hline
KM3NeT-ARCA & Mean $\delta$ & 2$\times 10^{-1}$ \\ 
\hline
IceCube & $0^{\circ}<\delta<90^{\circ}$ & 1$\times 10^{-1}$ \\ 
& $0^{\circ}<\delta<30^{\circ}$ & 2$\times 10^{-1}$ \\
\hline
\end{tabular}
\caption{Expected number of muon neutrino events in several detectors from an individual choked GRB with $L_{\rm iso}=10^{50}$~erg/s, t$_{\rm jet}=1000$~s, $E_{\rm{p,\,jet}}= 10^{53}$~erg in the IS frame and $\Gamma=100$ located at different declination bands. Effective areas for each detector were taken from \cite{ANTARES, KM3, IceCube}.}
\label{tab:1}
\end{table}


\section{Diffuse neutrinos from choked GRBs}
\label{sec:5}
To estimate the diffuse neutrino spectrum, we assume that the rate of choked jets $R(z)$ at redshift $z$ follows the star formation rate even if there are indications that the evolution of normal GRBs may deviate from the star formation rate $\rho(z)$ ~\citep{Lloyd,Petrosian},
i.e.,
\begin{equation}\label{Rcj}
R(z)=R_0\rho(z),
\end{equation}
where the star formation rate is ~\citep{madau2014}
\begin{equation}
\rho(z)=\frac{(1+z)^{2.7}}{1+[(1+z)/2.9]^{5.6}}\, ,
\end{equation}
and $R_0 $ is the local rate of choked GRBs in $\rm{Gpc}^{-3} \, \rm{yr}^{-1}$ . 
In this paper we do not consider a luminosity function for choked GRBs, but we assume a constant luminosity. A more detailed study on the redshift distribution and luminosity function effects on the diffuse neutrino flux is out of the aim of the present paper but it will be a subject for future works.

\begin{figure}
    \centering
    \includegraphics[scale=0.4]{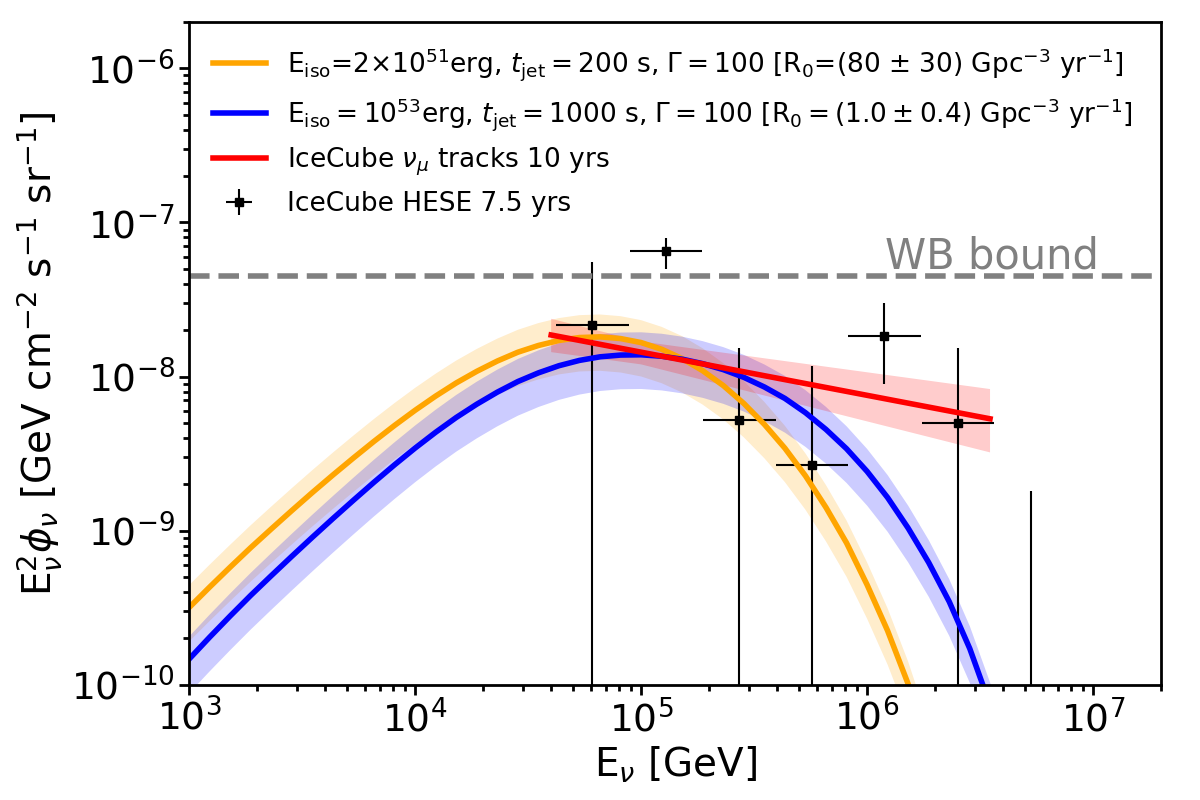}
    \caption{Diffuse neutrino spectra from choked GRBs expected on Earth, compared with the IceCube results on the HESE data (black squares) \cite{icecube2020} and on $\nu_{\mu}$ tracks (red line) \cite{stettner2019}. The blue line refers to the GRB simulation with $L_{\rm iso}=10^{50}$~erg/s, $t_{\rm jet}=10^3$~s and $\Gamma=100$ that is in better agreement with IceCube data as a function of $R_0$. The yellow line refers to the simulation with $L_{\rm iso}=10^{49}$~erg/s, $t_{\rm jet}=200$~s and $\Gamma=100$, again in better agreement with the observed cosmic neutrino flux as a function of $R_0$. Note that the inferred local rate of choked GRBs is consistent with the multiplet constraint from hard transient sources \cite{IC2019PRL} only in the less energetic case. In turn, a model with local rate of 1~Gpc$^{-3}$ yr$^{-1}$ is ruled out by the same constraints \cite{esmaili2018}. Additionally, the dashed grey line shows the Waxmann \& Bahcall upper bound \cite{waxbah2000}.}
    \label{fig:5}
\end{figure}

The diffuse neutrino flux is calculated via integrating the neutrino spectrum over the redshift from 0 to 8, i.e. ~\citep{murase2007}:
\begin{equation}
E_{\nu_{\mu}}^{{\rm obs}} \phi_{\nu_{\mu}}(E_{\nu_{\mu}}^{{\rm obs}}) = \frac{c}{4\pi H_0}\int_0^8\,E_{\nu_{\mu}}\frac{dN_{\nu_{\mu}}}{dE_{\nu_{\mu}}}((1+z)E^{{\rm obs}}_{\nu_{\mu}}) \frac{\frac{\Omega}{4\pi}R_0\rho(z)dz}{(1+z)\sqrt{\Omega_{\Lambda}+\Omega_{\rm M}(1+z)^3}}
\, ,
\end{equation}
where the cosmological parameters are adopted as
$H_0=70~{\rm km \, s^{-1} \, Mpc^{-1}}$, $\Omega_{\rm M}=0.3$, and $\Omega_{\Lambda}=0.7$,
$dN_{\nu_{\mu}}(E_{\nu_{\mu}})/dE_{\nu_{\mu}}$ is the differential neutrino spectrum at the source,
while $\Omega$ is the solid angle of the jet, $\Omega=2\pi(1-\cos\alpha)$, and $\alpha=0.2$~rad~$=11.5^{\circ}$ is its aperture.\\
In order to obtain the best value for the rate of choked GRBs, we compared the predicted flux of neutrino on Earth with the diffuse neutrino flux measured by IceCube. As experimental data we used the results obtained by IceCube both with the analysis of HESE events \cite{icecube2020} and the parametrization of the astrophysical diffuse muon neutrino flux \cite{stettner2019}.
The flux predicted with the simulation was evaluated for several values of $R_0$: the best agreement between data and our model has been obtained for $R_0 = (1.0 \pm 0.4)$~Gpc$^{-3}$ yr$^{-1}$ (with a reduced $\chi^2 \simeq 2.5$). Fig.~\ref{fig:5} shows this comparison: the central blue line represents the flux foreseen for the selected value of $R_0$, while the blue band represents the 68$\%$ confidence level uncertainty on $R_0$.\\
Recent IceCube results \cite{IC2019PRL} indicate that transient sources characterized by hard spectra ($dN_\nu/dE_\nu \propto E_\nu^{-2.13}$), with a local rate larger than $\sim 50$ Gpc$^{-3}$yr$^{-1}$, might possibly originate the observed diffuse neutrino flux. Within our model, this constraint implies that the energy converted into proton kinetic energy should be smaller than few times $10^{51}$~erg, though this constraint is quite sensitive to the assumed spectral shape of neutrinos and should be hence considered with caution. To test the case of a less energetic choked GRB population being able to explain the observed cosmic neutrino flux, we also performed a simulation characterized by the following values of model parameters: $L_{\rm iso}=10^{49}$~erg/s, $t_{\rm jet}=200$~s, $\Gamma=100$, $t_{\rm var}=0.005$~s and $E_{\rm{p,\,jet}}=2\times 10^{51}$ erg in the IS frame. Such a realization is consistent with the model presented in Sec.~\ref{sec:1}, though it differs in the maximum energy of accelerated protons, which now reads $E^{\rm IS}_{\rm cut} \simeq 25$~TeV, and in the maximum energy of muons, amounting to $E_\mu^{\rm IS} \simeq 50$~TeV because of synchrotron cooling. As a result, we obtained the yellow curve shown in Fig.~\ref{fig:5} for the diffuse neutrino flux contribution emerging from such a source population, which is at the level of the IceCube data provided a local rate equal to $R_0 = (80 \pm 30)$~Gpc$^{-3}$ yr$^{-1}$. The latter result is also in agreement with the transient analysis multiplet constraint for hard spectra sources \cite{IC2019PRL}.\\

\section{Results and Discussion}
\label{sec:6}
Recent stacking analyses performed by the current neutrino telescopes lead to the conclusion that classical GRBs are not the main sources of the diffuse astrophysical neutrino flux detected by IceCube \cite{icecubeGRB,antaresGRB}. In addition, also targeted searches for neutrinos in spatial and timing coincidence with the prompt emission of those GRBs reported evidence for a lack of correlation in neutrino data \cite{antaresBrightGRB}. Such results have increased even more the interest in an alternative GRB class opaque to gamma rays, choked GRBs, which would also be consistent with the IGRB observed by Fermi \cite{fermi2015}.\\We have estimated the neutrino fluxes from individual choked GRBs and the number of expected events for present and future neutrino telescopes. This result was obtained for the following set of benchmark parameters: $z=1$, $L_{\rm{iso}}=10^{50}$~erg/s, $\Gamma=100$, $t_{\rm jet}=1000$~s and assuming that $E_{\rm p,jet}=10^{53}$~erg in the IS frame.
We find that, for the declination in which each detector performs better, the number of muon neutrino events expected from an individual choked GRB source is equal to $5 \times 10^{-3}$ for ANTARES, and $2 \times 10^{-1}$ for IceCube and KM3NeT-ARCA (see Table \ref{tab:1}). These estimates scale almost linearly with the amount of energy channeled into protons. The number of neutrino events expected from each of these sources is in agreement with the lack of spatial anisotropies in the sky map distribution of IceCube neutrinos. Moreover, the lack of association of the latter events with known electromagnetic counterparts is to date consistent with an extra-galactic origin of such neutrinos from photon-dark sources, like choked GRBs.\\
We have also estimated the diffuse neutrino flux from the population of choked GRBs and compared it to the IceCube latest results \cite{icecube2020,stettner2019}. 
Through a best fit procedure we found that, if the energy that goes into accelerated protons is $E_{\rm p,jet} \sim 10^{51}-10^{53}$~erg in the IS frame, then IceCube data are reproduced for a local rate of choked GRBs of about $80-1$~Gpc$^{-3}$yr$^{-1}$, respectively. Note that, if $E_{\rm p,jet}$ is decreased, the local rate of choked GRBs has to increase in order to reproduce the IceCube data: this is consistent with what found in the literature by other authors \citep{MuraseWaxman,Senno}. \\
In order to associate the origin of the IceCube data to choked GRBs, the \textquoteleft\textquoteleft smoking gun" would be the detection of the latter at electromagnetic bands different than gamma rays. Preliminary studies have shown that these sources may emit in the X-ray, optical and UV band \citep{Irwin,Perna,Zhu}. We plan to extend this work and explore the possibility to perform a follow up of choked neutrino sources in these bands.

\acknowledgments
The authors thank the anonymous referee for the constructive comments, which significantly improved the manuscript. AC, SC, IDP and AZ acknowledge the support from Sapienza Università di Roma through the grant ID RM120172AEF49A82.

\newpage
\appendix

\section{Appendix - Analytic calculations for photo-meson interactions of protons with thermal photons}\label{app:a}
To develop the kinematics relations used in the code, we consider three reference systems: the source co-moving reference system, the IS reference system, and the center of mass system (these kinematic quantities are respectively indicated: plain, with the superscript ${\scriptstyle\text{IS}}$, with an asterisk \textbf{*}). We use natural units where $\hbar $ = c = 1, and we indicate with $\beta_a$ the speed of a generic particle ($\gamma_a=1/\sqrt{1-\beta_a^2}$), with $E_a$ its energy and with $m_a$ its mass. The invariant energy of the system, $\sqrt{s}$, is conserved and defined by the energy-momentum four-vector:
\begin{equation}\nonumber
s=(E_{\gamma}+E_p)^2 - (p_{\gamma}+p_p)^2=
\end{equation}
\begin{equation}\nonumber
=E_p^2 +E_{\gamma}^2 +2E_pE_{\gamma}-p_p^2 -E_{\gamma}^2-2p_pE_{\gamma}\cos\theta=
\end{equation}
\begin{equation}\label{eq:B1}
=m_p^2+2E_pE_{\gamma}(1-\cos\theta)
\,
.
\end{equation}
This relation holds in all reference frames. The minimum energy for a proton in the IS frame to produce a $\Delta^+$ resonance when interacting with a photon of energy $E^{\rm{IS}}_{\gamma}$ is shown in Eq.~\eqref{eq:2_4}: this is the threshold condition to produce the $\Delta^+$ resonance. 

The center of mass Lorentz factor is given by:
\begin{equation}\label{eq:B2}
\gamma^*=\frac{E^{\rm{IS}}_{p}+E^{\rm{IS}}_{\gamma}}{\sqrt{s}}\simeq \frac{E^{\rm{IS}}_{p}}{\sqrt{s}}
\,
.
\end{equation}
Given the enormous energies of the protons with respect to the photons, the angle between the direction along which the Lorentz boost of the center of mass is considered and the propagation direction of the proton is negligible. For simplicity, we will consider the Lorentz boost along the direction of observation. 

To get the particles spectra in the source co-moving frame, one needs to boost the energies in the IS by the jet's Lorentz factor, $\Gamma$.
In this reference system the following relation holds:
\begin{equation}
\label{eq:B3}
E_{p}=\Gamma\, E^{\rm{IS}}_{p}.
\end{equation}

Note that these quantities have not yet been corrected for the redshift due to cosmological distances.

\subsection{Delta Resonance Region}
\begin{equation}
\label{eq:B4}
p+\gamma \to \Delta^+ \to \begin{cases} n + \pi^+ \\ p + \pi^0\end{cases} \,
.
\end{equation}
In this region the invariant energy of the system $\sqrt{s}$ is fixed equal to the $\Delta^+$ resonance mass. Once formed, the resonance decays at rest in the center of mass system, and since the Lorentz boost is enormous, the decay products in the IS system continue along the same direction of the incident proton. At the resonance, the simulation code, takes into account the Clebsch-Gordon coefficients, and the two decay channels are produced with the following ratio:
\begin{equation}
\label{eq:B5}
\frac{\mathrm{BR}(p\gamma \to p\pi^0)}{\mathrm{BR}(p\gamma \to n\pi^+)}=\frac{2/3}{1/3}=2
\,
.
\end{equation}


By solving the interaction kinematics in this energy region (which corresponds to $0.2\, \mathrm{GeV}\le\epsilon_r<0.5\, \mathrm{GeV}$), we evaluate the energies of the secondary particles in the center of mass frame: 
\begin{equation}\label{eq:B6}
E_{\pi}^*=\sqrt{(p_{\pi}^*)^2 + m_{\pi}^2}=\frac{s+m_{\pi}^2-m_N^2}{2\sqrt{s}}
\,
,
\end{equation}
\begin{equation}\label{eq:B7}
E_N^*=\frac{s-m_{\pi}^2+m_N^2}{2\sqrt{s}}
\,
,
\end{equation} 
N being the nucleon produced together with the pion (proton or neutron).

In order to calculate the pion energy in the IS reference frame we must apply the Lorentz transformations that are given by (considering the IS moving towards the center of mass system along the boost direction, the x axis, with speed -$\beta^*$):
\begin{equation}
\label{eq:B8}
\begin{cases} 
p^{\rm{IS}}_x=\gamma(p_x^*+\beta^* E^*)\\p^{\rm{IS}}_y=p_y^* \\ 
p^{\rm{IS}}_z=p_z^* \\ 
E^{\rm{IS}}=\gamma(E^*+\beta^* p_x^*)
\end{cases}
\,
,
\end{equation}
where the speed and the Lorentz factor of the resonance can be written as $\beta^*\simeq 1$ and $\gamma=\gamma^*$. Considering that the momentum of the produced particles in a two-body decay is equal to:
\begin{equation}
\label{eq:B9}
p_N^*=p_{\pi}^*=\frac{\sqrt{s^2-2s(m_{\pi}^2+m_N^2)+(m_{\pi}^2-m_N^2)^2}}{2\sqrt{s}}
\,
,
\end{equation}
we finally get the pion energy in the co-moving reference frame ($\beta_{\pi}=1$):
\begin{equation}
\label{eq:B10}
E_{\pi}=\Gamma\gamma^*(E_{\pi}^*+p_x^*)=\frac{E_p}{\sqrt{s}}(E_{\pi}^*+p^*\cos\phi)
\,
,
\end{equation}
where $p^*$ is the common value of $p_N^*=p_{\pi}^*$, and $0^{\circ}\le\phi\le180^{\circ}$ is the angle, with respect to the proton direction, at which the pion is emitted from the $\Delta^+$ decay in its center of mass reference system.

\subsection{Charged Pion Decays}
\begin{equation}
\label{eq:B11}
{\pi}^+ \to \mu^+ + \nu_{\mu}
\,
,
\end{equation}
\begin{equation}
\label{eq:B12}
{\pi}^- \to \mu^- + \bar{\nu}_{\mu}
\,
.
\end{equation}
We evaluate the kinematics of the $\pi^+$ decay, considering that the same relations will hold in the $\pi^-$ decay, which will emerge only in the out of resonance region (see Sections \ref{A:5}, \ref{A:6}).
The muon and the neutrino energies in the $\pi^{+}$ center of mass system (Equation \ref{eq:B11}) are:
\begin{equation}
\label{eq:B13}
E_{\mu^+}^*=\frac{m_{\pi^+}^2 + m_{\mu^+}^2}{2m_{\pi^+}},
\end{equation}
\begin{equation}
\label{eq:B14}
E_{\nu_{\mu}}^*=\frac{m_{\pi^+}^2 - m_{\mu^+}^2}{2m_{\pi^+}}.
\end{equation}
For the neutrino, $p_{\nu}=E_{\nu}$, so the momentum of the decay products is:
\begin{equation}
\label{eq:B15}
p_{\nu_{\mu}}^*=p_{\mu^+}^*=\frac{m_{\pi^+}^2 - m_{\mu^+}^2}{2m_{\pi^+}}
\,
.
\end{equation} 

Using Lorentz transformations, we get the energies in the co-moving reference frame, using the charged pion Lorentz factor $\gamma_{\pi^+}=E_{\pi^+}/m_{\pi^+}$:
\begin{equation}
\label{eq:B16}
E_{\mu^+}=\gamma_{\pi^+}(E_{\mu^+}^*+p_{\mu^+}^*\cos\delta),
\end{equation}
\begin{equation}
\label{eq:B17}
E_{\nu_{\mu}}=\gamma_{\pi^+}(E_{\nu_{\mu}}^*-p_{\nu_{\mu}}^*\cos\delta),
\end{equation}
where $\delta$ is the angle between the muon and the decaying pion, $0^{\circ}\le\delta\le180^{\circ}$. 

\subsection{Neutral Pion Decay}
\begin{equation}
\label{eq:B18}
{\pi}^0 \to \gamma + \gamma\,
.
\end{equation}
In the reference frame in which the pion is at rest, one has:
\begin{equation}
\label{eq:B19}
p_{\gamma}^*=E_{\gamma}^*=\frac{1}{2}m_{\pi^0}.
\end{equation}
So the energies of the photons resulting from this decay process are:
\begin{equation}
\label{eq:B20}
E_{\gamma}=\gamma_{\pi^0}(E_{\gamma}^*\pm p_{\gamma}^*\cos\zeta)=\frac{E_{\pi^0}}{2}(1\pm \cos\zeta),
\end{equation} 
where $\gamma_{\pi^0}=E_{\pi^0}/m_{\pi^0}$ (and the decay products are emitted with an angle $0^{\circ}\le\zeta\le180^{\circ}$). The $\pm$ sign refers to the opposite directions of emission of the two photons.

\subsection{Muon Decay}
\begin{equation}
\label{eq:B21}
\mu^- \to e^- +\bar{\nu}_e+\nu_{\mu}\,
,
\end{equation}
\begin{equation}\label{eq:B22}
\mu^+ \to e^+ +\nu_e+ \bar{\nu}_{\mu}\,
.
\end{equation}
In the following we refer with $\nu_{\mu}$ and $\nu_e$ to indicate muon and electron neutrinos and antineutrinos. The high-energy neutrino flux reaching the detectors includes both the $\nu_{\mu}$ produced by the charged pion decay (see Eq.~\eqref{eq:B12}), and the $\nu_{e}$ and $\nu_{\mu}$ produced by the decay of the muon.

As a consequence of the non-conservation of parity, muons coming from pion decays are strongly polarized: those coming from the $\pi^{\pm}$ decay have a right-handed helicity. In the system in which the muon is at rest, the particle distribution as a function of energy and of the emission angle $\omega$ is given by:
\begin{equation}\label{eq:B23}
\frac{\mathrm{d}N}{\mathrm{d}x\mathrm{d}\omega}=\frac{1}{4\pi}(f_0(x)-f_1(x)\cos\omega)\,
,
\end{equation}
where $x=2E_l^*/m_{\mu}$, the Bjorken x, is the fraction of the available energy ($m_{\mu}$) carried by the lepton \textit{l} (\textit{l} = $e^-,\, \bar{\nu}_e,\, \nu_{\mu}$), thus: $0\le x \le1$. The variable $\omega$ defines the angle between the daughter particle and the muon spin. The functions appearing in the above equation can be obtained through the Michel parameters, which describe the phase space distribution of leptonic decays of charged leptons. With these parameters, one can obtain:
\begin{itemize}
\item for the muon neutrino
\begin{equation}
\label{eq:B24}
\begin{cases} 
f_0(x)=2x^2(3-2x) \\ 
f_1(x)=2x^2(1-2x)\end{cases}
\,
;
\end{equation}
\item for the electron neutrino
\begin{equation}
\label{eq:B25}
\begin{cases} 
f_0(x)=12x^2(1-x) \\ 
f_1(x)=12x^2(1-2x)
\end{cases}
\,
.
\end{equation}
\end{itemize}
In the co-moving reference system, Equation~\eqref{eq:B23} becomes:
\begin{equation}
\label{eq:B26}
\frac{dN}{dy}=\frac{1}{\beta_{\mu}}(g_0(y,\beta_{\mu})-P_{\mu}g_1(y,\beta_{\mu}))
\,
,
\end{equation}
where 
\begin{equation}
\label{eq:B27}
y\simeq \frac{E_l}{E_{\mu}}
\,
,
\end{equation}
is the Bjorken y, namely the fraction of energy carried by the lepton \textit{l} in the co-moving frame, and $P_{\mu}$ is the muon spin projection along the direction of motion in the co-moving frame. $P_{\mu}$ is defined as ($m_{\pi}$ is the charged pion mass):
\begin{equation}
\label{eq:B28}
P_{\mu}=\frac{1}{\beta_{\mu}}\left(\frac{2E_{\pi}(m_{\mu}/m_{\pi})^2}{E_{\mu}(1-(m_{\mu}/m_{\pi})^2)}-\frac{1+(m_{\mu}/m_{\pi})^2)}{1-(m_{\mu}/m_{\pi})^2)}\right).
\end{equation}
Considering that in our case $\beta_{\mu}\simeq 1$, we have:
\begin{itemize}
\item for the muon neutrino 
\begin{equation}
\label{eq:B29}
\begin{cases} g_0(y)=\frac{5}{3}-3y^2+\frac{4}{3}y^3 \\ g_1(y)=\frac{1}{3}-3y^2+\frac{8}{3}y^3\end{cases}
\,
;
\end{equation}
\item for the electron neutrino
\begin{equation}
\label{eq:B30}
\begin{cases}g_0(y)=2-6y^2+4y^3 \\ g_1(y)=-2+12y-18y^2+8y^3 \end{cases}
\,
.
\end{equation}
\end{itemize}
The energy of each neutrino in the co-moving reference frame has been obtained by multiplying the muon energy defined by Eq.~\eqref{eq:B16} with a value \textit{y} extracted according to Equation~\eqref{eq:B26}.

This procedure is valid regardless of the decaying muon's charge, and it was used to obtain the neutrino energies coming from the decays shown in Eq.~\eqref{eq:B21} and Eq.~\eqref{eq:B22}.

\subsection{Delta Resonance Secondary Peak Region} \label{A:5}
In the photon energy range $0.5\, \mathrm{GeV}\le\epsilon_r<2\, \mathrm{GeV}$, the $\Delta^+$ is accompanied by a pion:
\begin{equation}
\label{eq:B31}
p + \gamma \to \Delta^+ + \pi^{\mathrm{A}}\,
.
\end{equation}
The invariant energy of the system is no longer equal to the mass of the $\Delta^+$ resonance, but it's given by Eq.~\eqref{eq:B1}. In the center of mass frame, the following kinematic relations hold:
\begin{equation}
\label{eq:B32}
E^*_{\Delta^+}=\frac{s+m^2_{\Delta^+}-m^2_{\pi^{\rm{A}}}}{2\sqrt{s}}
\,
,
\end{equation}
\begin{equation}
\label{eq:B33}
E^*_{\pi^{\rm{A}}}=\frac{s+m^2_{\pi^{\rm{A}}}-m^2_{\Delta^+}}{2\sqrt{s}}
\,
.
\end{equation}
Moreover, $p^*_{\pi^{\rm{A}}}$ is given by Eq. \ref{eq:B9} with $\pi=\pi^{\rm{A}}$ and $N=\Delta$. The $\pi^{\rm{A}}$ energy in the co-moving frame can be written as ($0^{\circ}\le\eta\le180^{\circ}$):
\begin{equation}
\label{eq:B34}
E_{\pi^{\rm{A}}}=\frac{E_p}{\sqrt{s}}(E^*_{\pi^{\rm{A}}}+p^*_{\pi^{\rm{A}}}\cos\eta)
\,
.
\end{equation}
The charged and neutral pions are produced with the following branching ratios:
\begin{itemize}
\item $1/15$ of the $A$ pions are neutral, $\pi^0$, which decay in two photons (Eq.~\eqref{eq:B18}) with energies given by Eq.~\eqref{eq:B20};
\item $8/15$ of the $A$ pions are positive charged, $\pi^+$, which decay in a muon neutrino and a positive muon (Eq.~\eqref{eq:B11}, which in turn decays in a positron, an electron neutrino and a muon antineutrino, Eq.~\eqref{eq:B22}). Energies for these particles are given by Eqs.~\eqref{eq:B16}, \eqref{eq:B17}, and \eqref{eq:B27};
\item $2/5$ of the $A$ pions are negative charged, $\pi^-$, which decay in a muon antineutrino and a muon (Eq.~\eqref{eq:B12}, which in turn decays in an electron, an electron antineutrino and a muon neutrino, Eq.~\eqref{eq:B21}). Energies for these particles are also given by Eqs.~\eqref{eq:B16}, \eqref{eq:B17}, and \eqref{eq:B27}.
\end{itemize}

In order to consider all the processes that can originate neutrinos and photons, we have to take into account the $\Delta^+$ decay: 
\begin{equation}
\label{eq:B35}
\Delta^+ \to \mathcal{N} + \pi^{\mathrm{B}}\,
.
\end{equation}
In the $\Delta^+$ resonance rest frame, we have:
\begin{equation}
\label{eq:B36}
|p^*_{\pi^{\mathrm{B}}}|=|p^*_{\mathcal{N}}|=
\frac{\sqrt{[(m^2_{\Delta^+}-(m_{\mathcal{N}}+m_{\pi^{\mathrm{B}}})^2)(m^2_{\Delta^+}-(m_{\mathcal{N}}-m_{\pi^{\mathrm{B}}})^2)]}}{2m_{\Delta^+}}
\,
,
\end{equation}
\begin{equation}
\label{eq:B37}
E^*_{\pi^{\mathrm{B}}}=\frac{s+m^2_{\pi^{\rm{B}}}-m^2_{\mathcal{N}}}{2m_{\Delta^+}}
\,
.
\end{equation}
Finally, the $B$ pion energy in the co-moving frame is given by:
\begin{equation}
\label{eq:B38}
E_{\pi^{\mathrm{B}}}=\frac{E_{\Delta^+}}{m_{\Delta^+}}(E_{\pi^{\mathrm{B}}}^*+p^*_{\pi^{\mathrm{B}}}\cos\rho)
\,
,
\end{equation}
where $0^{\circ}\le\rho\le180^{\circ}$ is the emitting angle of the pion, and $E_{\Delta^+}\sim E_p-E_{\pi^{\mathrm{A}}}$. 
The code extracts randomly the branching ratios for $\pi^{\mathrm{B}}$, which are: 
\begin{itemize}
\item $18/45$ of $B$ pions are neutral pions;
\item $19/45$ of $B$ pions are positive charged pions;
\item $8/45$ of $B$ pions are negative charged pions.
\end{itemize}
The energies of the particles emerging from the pion decay are obtained as already described in the above sections.

\subsection{Multipion Production Region} \label{A:6}
When the photon energy exceeds 2~GeV, most of the energy lost by the proton ($\sim 0.6\,E_p$) is split equally among three pions, and the neutral and charged pions are approximately produced in equal numbers (this is a simplified treatment, as found in e.g. \citep{atoyan2003, kelner2008, hummer2010}). So the single pion energy is equal to $E_{\pi}=0.2\,E_p$, and the particles resulting from the pion decay are simulated according with the procedures described in the previous sections.


\end{document}